# Sensemaking Through Making: Developing Clinical Domain Knowledge by Crafting Synthetic Datasets and Prototyping System Architectures

Mihnea Stefan Calota[*]
m.s.calota1@tue.nl
Eindhoven University of Technology
Eindhoven, The Netherlands

Wessel Nieuwenhuys[*]
w.w.nieuwenhuys@tue.nl
Eindhoven University of Technology
Eindhoven, The Netherlands

Janet Yi-Ching Huang
y.c.huang@tue.nl
Eindhoven University of Technology
Eindhoven, The Netherlands

Lin-Lin Chen
l.chen@tue.nl
Eindhoven University of Technology
Eindhoven, The Netherlands

Mathias Funk
m.funk@tue.nl
Eindhoven University of Technology
Eindhoven, The Netherlands

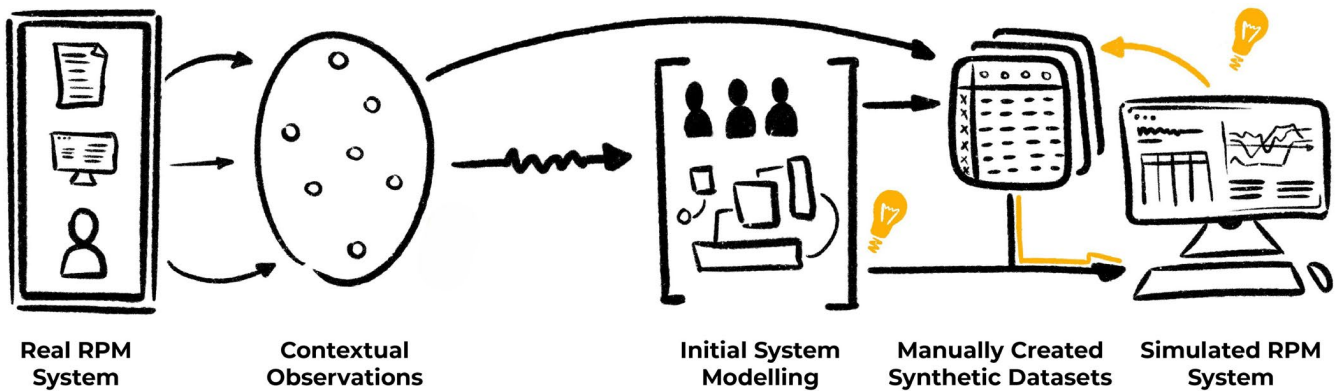

Figure 1: We present an alternative approach for designers to develop a deeper understanding of how a complex healthcare system works. The *Remote Patient Monitoring (RPM) system* is a closed, mostly opaque box, that is difficult to access. *Contextual observations* of the real system are used to create a proxy *system model*. The modeled system architecture and the observations are used to *manually craft synthetic datasets* that can populate the system model. Lastly, the designer continuously *reflects* (yellow) on the created data, and iteratively builds a *simulated RPM system*, based on those reflections. The process of *building, reflecting and iterating* allows the designer to learn and quickly develop domain knowledge that can be used when further engaging with clinical stakeholders.

## Abstract

Designers have ample opportunities to impact the healthcare domain. However, hospitals are often closed ecosystems that pose challenges in engaging clinical stakeholders, developing domain knowledge, and accessing relevant systems and data. In this paper, we introduce a making-oriented approach to help designers understand the intricacies of their target healthcare context. Using Remote Patient Monitoring (RPM) as a case study, we explore how manually crafting synthetic datasets based on real-world observations enables designers to learn about complex data-driven healthcare systems. Our process involves observing and modeling the real-world RPM context, crafting synthetic datasets, and iteratively prototyping a simplified RPM system that balances contextual richness and intentional abstraction. Through this iterative process of sensemaking through making, designers can still develop context familiarity, when direct access to the actual healthcare system is limited. Our approach emphasizes the value of hands-on interaction with data structures to support designers in understanding opaque healthcare systems.

---

[*]Both authors contributed equally to this research.

## 1 Introduction

Healthcare has always been a domain with a high inherent complexity, and over the years it has become increasingly data-driven, utilizing clinical and contextual patient data [3]. By leveraging this data, clinicians can tailor care programs to individual needs and employ predictive analysis to intervene early, reducing hospital (re-)admissions [15]. As these technologies evolve, designers have the opportunity to support their implementation, ensuring that technological advancements can be effectively integrated into the healthcare system [26].

Designers, despite this potential for impact, often face difficulties working in healthcare settings. Engaging clinical stakeholders, deeply understanding the context and the complex systems they are designing for, and gaining access to relevant hospital data remain significant challenges [4, 14]. Previous studies highlight how restricted access to the medical domain limits designers' ability to thoroughly explore and develop domain knowledge [14]. This constraint is crucial, as a *poor contextual fit* is one of the leading causes of failure when technologies transition from the design studio to real-world implementation [23, 27]. Even when access is granted, it is often fragmented, providing only a partial view of the system and making it difficult to design holistic, integrated solutions.

Traditionally, ethnographic methods such as interviews and observations have helped designers understand their target contexts [5]. However, these approaches often provide only limited access to hospital workflows and are highly dependent on the clinicians' availability, limiting their usefulness in developing a complete understanding of system dependencies. Furthermore, newer methodologies like Data-Enabled Design (DED) [6, 7] incorporate data as a material for iterative design, facilitating ongoing cycles of reflection and refinement. Yet, in clinical settings, the constraints of data privacy, time, and access make it impractical to conduct DED [21]. As previous studies explored [21], the clinical domain does not afford designers the opportunity for lengthy and in-depth explorations to develop a good understanding of the domain; designers need to accumulate that knowledge upfront. However, the clinicians have busy agendas, and the data and systems are too sensitive to be easily accessed for exploratory reasons.

Facing these constraints, we propose giving designers a wider repertoire of tools they can use to prepare before engaging with the clinical domain and develop prior knowledge about their target medical context, even when direct access to the domain is limited. Design researchers have long recognized the role of prototyping as an inquiry tool, where designers engage with material artifacts to generate new knowledge [25]. Thus, for designers, sensemaking and making are deeply intertwined [11]. Sensemaking focuses on how humans transform ambiguous and complex situations into understandable and actionable information [17, 19, 22].

As such, we integrate existing sensemaking and prototyping approaches to address the unique challenges of healthcare design, arguing that designers can use data to iteratively build, reflect, and learn about complex medical systems. We contribute an approach that prioritizes hands-on interaction with data structures, where designers prototype and reflect on synthetic datasets and a simulated medical system, to increase their practical understanding of the inner workings and intricacies of a domain (**Figure 1**).

## 2 An alternative approach for developing domain knowledge about an RPM system

In this work, we ask *how can designers develop domain knowledge when access to the target systems, users and stakeholders is limited?* As a case study, we look at Remote Patient Monitoring (RPM) for heart failure, a subset of data-driven healthcare, which focuses on providing care at home for patients with chronic diseases. [1]. By analyzing patient-measured data, clinicians have the ability to quickly react to changes in patient status without the need for consultations, saving in-person time and costs [10, 20]. However, adopting RPM requires healthcare professionals (HCPs) to handle increasingly large volumes of data and determine appropriate responses, thus expanding their data work responsibilities [13, 24]. As the global elderly population grows, the reliance on RPM systems and the corresponding strain on HCPs are expected to increase [16].

By *sensemaking through making* synthetic datasets and simulating healthcare system interactions, designers can actively construct their understanding of hidden system rules, workflows, and actor behaviors in clinical domain. Grounded in the Research through Design framework, this approach embraces active learning through making and reflection [12, 18], through several steps (**Figure 1**).

(1) *Observing* real-world RPM settings through ethnographic methods, such as interviews and direct observations.
(2) *Modeling* the RPM system by defining key actors, entities and relationships, and the system architecture.
(3) *Crafting and reflecting* upon synthetic datasets representative of this model, informed by contextual observations.
(4) *Building* a simplified, synthetic RPM system to facilitate interaction with and visualization of data.

We manually created synthetic datasets for a sample of 10 patients and 6 HCPs, covering a period of 6 months. We hypothesized that the resulting simulated system would serve as a tool for context inquiry and domain knowledge generation. As such, we prioritized qualitative richness over the statistical significance of the dataset. Moreover, using the lens that designers *learn by doing*, we envisioned that participating in the process is what would provide a deeper understanding of the system's behavior, compared to simply exploring the resulting datasets.

### 2.1 Observing RPM in the wild

We started by developing a familiarity with the context of RPM for heart failure through interviews and direct observations [8]. We also made contextual observations from real EHRs and an RPM application (i.e., Luscii [2]) used by a hospital in the Netherlands.

While we lacked direct access to the hospital's datasets, we were shown selected examples on-site, though without the ability to interact with or manipulate the data. These observations gave us an initial understanding of the type of data workflows HCPs employ and highlighted two important factors: (1) the system is more "messy" in reality than expected, with both structured and unstructured data types, as well as many duplicate information, and (2) many parts of the system remain opaque. We considered this real-life "messiness" an essential feature to replicate in our synthetic model, as it would add qualitative richness and mirror the complexities designers need to account for. The opaqueness of the system is what motivated us to look for alternative ways to engage with it, and what led to the next steps in our method (**Figure 1**).

## 2.2 Turning observations into guidelines for system modeling

The goal of this step was to extract insights from our field observations to decide which parts of the observable system could already be modeled and which rules still needed to emerge from our future steps. To model the system, we defined its actors, components, and the logic and relationships linking them.

*Entities and Relationships.* Identifying entities and their relationships was directly informed by the observed EHR and RPM datasets. We established six entities encapsulating most of the interactions one patient has within the RPM system: patients, HCPs, health measurements (e.g., blood pressure, heart rate, weight, open comments), alerts for abnormal measurements, medication changes, and hospital admissions.

*System Logic.* Defining the system logic proved more challenging, as field ethnography alone did not yield a nuanced understanding of the RPM system's dynamic rules and dependencies. While our context knowledge facilitated the development of entities and fundamental relationships between datasets, we encountered uncertainty in defining dynamic rules and dependencies.

## 2.3 Crafting synthetic datasets

Throughout the dataset construction stages, we engaged in a continuous loop of "building, reflecting, and learning", by creating data points, reflecting on how they might interact with other data, and distilling new insights about the synthetic system. The insights were continuously applied at multiple levels—at a high level, they informed broad categories of data (e.g., ensuring that consultations reflect the distinct writing style of different HCPs), and at a low level, they guided individual decisions (e.g., a call makes sense here since it is Friday and otherwise there will be no contact until Monday).

We learned early that data dependencies follow a logical order: actors must exist before they can act; patients first submit independent measurements, which trigger alerts; nurses then resolve these alerts, sometimes leading to medication changes or hospital admissions, while HCPs generate consultations based on all of these events. Some hidden dependencies—such as medication affecting measurements or a nurse's confidence influencing alert responses—emerged only through the iterative reflection that was prompted by the process. The following data types were prototyped.

*Patients and HCPs Data.* We created patient entities who would submit measurements at different intervals, exhibiting varying degrees of instability, receiving different levels of support at home, and managing different comorbidities. Similarly, the different HCPs had different personalities, documentation styles, and varying levels of experience with RPM.

*Patient Measurements Data.* The measurements of the patients were generated as a chronological series of values, incorporating a controlled level of randomness within set parameters. Patient characteristics influenced data variability; some patients exhibited stable data, while others experienced higher fluctuations or occasional spikes, leading to more frequent alerts.

To reflect real-world data patterns, open comments from patients were manually added in places where measurements suggested a potential comment might arise. Extra comments were incorporated to simulate potentially irrelevant data, aligning with data traces observed in the real system.

*Generating Alerts and Nurse Response Data.* Mirroring real-world practices, we automatically triggered alerts based on patient-specific thresholds and abrupt changes in the synthetic measurements.

Stepping into the shoes of the HCP workflow, we opened each alert manually, assessed the data, and decided on a response from the perspective of the HCP persona. While embodying the HCPs, we had fixed response options, as observed in the real context: call the patient, adjust a medication, contact a colleague, or dismiss the alert. Approximately 13% of measurements triggered alerts, which was in line with what we observed in reality.

*Patient Admissions Data.* Based on the patient personas, measurement trends, and nurse responses, we manually added admission periods to each patient's history. Reflecting on how these would show up in the data, we also realized that there would be a decrease in RPM system interaction during a hospitalization, and the HCP's contact with the patient would be modeled differently.

*Medication Changes Data.* We recorded medication changes in a separate dataset, as this fragmentation was observed in the real EHR. Drawing from one of our earlier learning points, we created medication change entries based on changes observed in the synthetic measurement trends and on generated nurse responses.

## 2.4 Building a simplified synthetic RPM system

Although the goal of this process was not to start designing interventions yet, we settled on creating a tool that improved our workflow, by facilitating data input, editing, and browsing through large amounts of data, over simply using tabular data editors (i.e., Excel). The app acted as a simplified RPM system that prioritized visualizing data points in graphs, summarizing patient histories, and showing connections between actors and events. The simulated RPM interface served as a sandbox to explore the first-person perspective of HCPs and experience potential painpoints or frustrations in their normal data workflows. Examples of the app's interface can be found in the **supplementary materials**.

## 3 Discussion

In this case study, we presented a designerly strategy for navigating restricted-access healthcare contexts, building on established sensemaking and prototyping approaches. Here, we expand on our learnings and the implications of this method for future designers.

### 3.1 Developing domain knowledge to prepare for design activities

This approach does not replace traditional methods such as ethnography or participatory design but serves as a complementary tool for early-stage domain exploration, especially when direct access to stakeholders or real data is limited. Through the inspection of data structures and observations of the real-life system and its context, followed by prototyping system representations, we partly engaged in *reverse engineering* the real-world system [9, 12]. In this process it was necessary to continuously reflect on the role of data as being rooted in our need to understand and articulate the hidden inter-dependencies, workflows, and structures of the target RPM system.

The presented approach extends familiar methodologies—breaking systems apart, reassembling them, and engaging in hands-on interaction to uncover how they function—while allowing designers to explore domains considered less accessible before, such as healthcare. In this process, the designer builds a much-needed context familiarity, and is prompted to find which questions are relevant to ask and where it is important to dig deeper, as we have shown through our example use case. The process hinges on a careful balance between embracing messiness and abstraction, which can help designers account for complexity and variability in their design processes, prompting them to think about concrete dependencies in the context they are designing for. Ultimately, this approach centers on making-oriented activities as a way of preparing for domain engagement, so that when the limited access to stakeholders or data structures is obtained, the designer can take full advantage of it.

### 3.2 Hands-on engagement with data leads to knowledge

It is well-documented that the act of crafting prototypes and research probes is a fundamental practice for designers wishing to explore a problem space [11, 18]. Similarly, we found that engaging with data in a hands-on process to construct a model of the hospital system had similar benefits for our understanding of RPM context. We started from a limited amount of assumptions and observations, and then we engaged in the act of crafting to resolve the tension between uncertainty and the need for concreteness.

The *sensemaking through making* approach encouraged us to step into the shoes of the HCPs and experience their cognitive processes in a concrete and tangible environment supported by the data structures and system interface we created. This way, we could go beyond speculation and inquire into the rationale of the actors involved in our target context while being grounded in a realistic setting. For example, as we opened and assessed each generated alert to create responses, we engaged with the HCP's position and mindset. After a prolonged period of acting as HCPs in the synthetic system, we developed new strategies for data interpretation and decision-making, which indicate an increased understanding of the domain's intricacies. For example, we initially based the simulated HCP response purely on a singular data point. However, as we developed a deeper understanding of the HCP perspective, our decision-making incorporated contextual factors (e.g., day of the week or inferred patient personality). Furthermore, new system logic emerged—something we struggled with in the earlier phases of data creation.

Recreating the tangible components of the RPM system by building a web interface further supported the value of *making to learn*. As isolating trends is difficult in tabular data, the visualizations we created in the app allowed us to view recent consultations, admissions, and patient history more easily. The process of constructing the data structures of our target system inadvertently prompted the construction of the system's interface as well. Moreover, the act of designing this simplified synthetic RPM system highlighted the challenges that HCPs face when working with the real RPM system. Cross-referencing events, patient history summaries, and an overview of previous consultations were all features we introduced to our app out of a practical need that we hypothesize is felt by the clinicians as well.

Initially, we thought that HCPs would need support interpreting alerts, but we didn't grasp how this process unfolds. Our exploration revealed that gathering data to make decisions is where friction might actually exist, as we felt the need for new interfaces that aggregate data into timelines and summaries. While initially we assumed there is no standardized procedure for deciding to call patients, we now suspect that specific, implied decision flowcharts guide practice (e.g., a weight spike is typically accompanied by a medication change). Lastly, we now see collaboration among HCPs as an area that might need support, something we did not consider initially. Overall, the process laid a strong foundation for engaging with clinicians in precise discussions that can take advantage of their limited availability.

### 3.3 Limitations

A core limitation of this approach is its heavy reliance on synthetic data and the designers' own reflections to generate domain knowledge. This can introduce biases and incorrect assumptions about real clinical workflows. While synthetic datasets help designers explore system dynamics, they cannot replace real patient data or the lived experiences of clinicians. That said, this approach is not intended to replicate reality accurately, but rather to help designers develop context understanding and identify concrete assumptions before engaging with stakeholders. The speculative nature of this approach, while a limitation for statistical accuracy, fosters active learning, supporting the early stages of design.

In our case, we were fortunate to explore datasets, observe nurse practices, and ask questions, even if this access was insufficient for in-depth exploration. It is important to note that this approach hinges on these glimpses into the field, as they are instrumental in creating a model that correlates with reality. This method does not function without a way to anchor the *sensemaking through making* process in the actual clinical context. Instead, it builds upon limited observations through informed speculations on data and system architectures, ensuring that insights remain grounded.

Designers should seize every opportunity to validate and refine their synthetic models in collaboration with clinicians, as this could lead to less influence of the designer's own biases, ensure real-world applicability and root the model in clinical practice. The existence of these models alone, however, already represents a step forward.

## 4 Conclusion

This study shows how *sensemaking through making* can help designers navigate restricted healthcare contexts by building domain understanding through making synthetic datasets and system prototyping. While not a replacement for real-world data, this approach helps designers ask better questions, uncover hidden system dynamics, and prepare for nuanced clinician engagement. We encourage future healthcare designers to try our approach in the field, integrate it into their design process, and reflect on its applicability and use.